\documentclass[9pt, conference]{IEEEtran}
\IEEEoverridecommandlockouts
\usepackage{cite}
\usepackage{amsmath,amssymb,amsfonts}
\usepackage{algorithmic,graphicx}
\usepackage{textcomp}
\usepackage{xcolor}
\usepackage{multirow,multicol}
\usepackage{hyperref}
\makeatletter

\def\BibTeX{{\rm B\kern-.05em{\sc i\kern-.025em b}\kern-.08em
    T\kern-.1667em\lower.7ex\hbox{E}\kern-.125emX}}
\begin{document}

\title{Metric Analysis for \\
Spatial Semantic Segmentation of Sound Scenes}
\author{
\IEEEauthorblockN{Mayank Mishra, Paul Magron, Romain Serizel}\\
\IEEEauthorblockA{Université de Lorraine, CNRS, Inria, LORIA\\
Nancy, France\\
Emails: mayank.mishra@loria.fr, paul.magron@inria.fr, romain.serizel@loria.fr}
}

\maketitle

\begin{abstract}
Spatial semantic segmentation of sound scenes (S5) consists of jointly performing audio source separation and sound event classification from a multichannel audio mixture. Evaluating S5 systems with separation and classification metrics individually makes system comparison difficult, whereas existing joint metrics, such as the class‑aware signal‑to‑distortion ratio (CA‑SDR), can conflate separation and labeling errors. In particular, CA-SDR relies on predicted class labels for source matching, which may obscure label swaps or misclassifications when the underlying source estimates remain perceptually correct. In this work, we introduce the class and source‑aware signal-to-distortion ratio (CASA‑SDR), a new metric that performs permutation‑invariant source matching before computing classification errors, thereby shifting from a classification-focused approach to a separation-focused approach. We first analyze CA‑SDR in controlled scenarios with oracle separation and synthetic classification errors, as well as under controlled cross‑contamination between sources, and compare its behavior to that of the classical SDR and CASA-SDR. We also study the impact of classification errors on the metrics by introducing error-based and source-based aggregation strategies. Finally, we compare CA-SDR and CASA-SDR on systems submitted to Task 4 of the DCASE 2025 challenge, highlighting the cases where CA‑SDR over‑penalizes label swaps or poorly separated sources, while CASA‑SDR provides a more interpretable separation‑centric assessment of S5 performance.
\end{abstract}


\section{Introduction}
\label{sec:intro}


The use of machine learning has become ubiquitous in today's daily life, as it has helped tackle more complex problems than ever and opened up opportunities for applications in technology aimed at specific needs. In particular, machine listening, which aims to design systems that automatically analyze audio scenes~\cite{virtanen2017computational}, has found novel applications in smart homes and smart cities~\cite{DING2024121902,VAFEIADIS2020103226, Gairí2025}.

Spatial semantic segmentation of sound scenes (S5) is a recent task that aims to jointly perform audio source separation and audio classification from a multi-channel audio input~\cite{nguyen2025,yasuda2025}. S5 is a step towards the development of audio systems that could help in performing machine listening, and is of particular importance in technologies to be used in home assisting living and immersive communications. Since S5 is a combination of two sub-tasks, one way to evaluate S5 systems is to evaluate audio source separation and classification separately by using respective metrics for each sub-task~\cite{Vincent2006bss, Liu_2023}. Another possibility is to consider one of the two as a primary task and evaluate the systems based on the metrics for that primary task. This is what was done to evaluate sound event detection and separation, a problem similar to S5 but where sound event detection was considered as a primary task over source separation~\cite{cornell2024dcase}. However, for certain applications, both tasks are equally important. For example, in immersive communications, the audio streams need to be properly separated before transmission, and they need to be properly classified in order to select the correct streams to transmit. The metric then needs to reflect both the separation and classification performance.

To that end, a class-aware signal-to-distortion ratio (CA-SDR) metric was proposed and used in the DCASE $2025$ challenge~\cite{nguyen2025}. CA-SDR utilizes a combination of an audio separation metric, the SDR \cite{makino2018audio, LeRoux2019sisdr} weighted by the true positives (TPs), false positives (FPs), and false negatives (FNs) obtained based on the class allocations done by the system. This approach puts a strong emphasis on classification, regardless of the similarity between the separated sources. As a result, it can tend to overlook potential mismatches between sources and their corresponding class labels. 

In this paper, we propose a new metric for evaluating S5 systems called the class and source-aware SDR (CASA-SDR). The metric takes into account the similarity between the predicted and reference sources before computing the classification errors. The goal of CASA-SDR is to provide an alternative that alleviates certain shortcomings of CA-SDR by focusing on a separation-centric approach to evaluation. Indeed, CASA-SDR first computes a permutation-invariant version of SDR and then checks for the classification errors, leading to a source-aware metric calculation. To analyze these metrics' behavior, we conduct experiments both in controlled conditions using a synthetic dataset, and on real-world systems that were submitted to Task 4 of the DCASE 2025 challenge. Our findings outline that while CA-SDR over-penalizes label swaps or poorly separated sources, CASA-SDR provides a more interpretable
separation-centric assessment of S5 performance.

The rest of this paper is structured as follows. First, in Section~\ref{sec:problem_statement} we introduce the S5 task, the metrics, and we underline possible separation and classification error scenarios. Then, we describe our experimental protocol in Section~\ref{sec:exp_protocol}. Section~\ref{sec:results_and_disc} presents and discusses experimental results. Finally, Section~\ref{sec:conc} concludes the paper.

\section{Evaluation Metrics For S5}
\label{sec:problem_statement}

\subsection{S5 definition and setup}
\label{sec:S5def}

The S5 task assumes the following setup~\cite{nguyen2025,yasuda2025}. We consider $M$ sound sources ($s_1, \dots, s_M$) that are distributed in a room. Among those sources, the first $N$ sources ($N<M$) are considered as target sources, and the remaining $N-M$ sources are considered as non-target sources (i.e., interferences).
Each source is associated with one sound event that is active at some point within the duration of the signal, and whose label is denoted $c_i$, such that $C = (c_1,\dots,c_N)$ denotes the set of true target class labels. For the non-target sources, the classes of the sound events are considered unknown.

The goal of an S5 system is to estimate each of the target sources that are present in the mixture, as well as their respective class labels. For the remainder of the paper, to simplify the analysis and align closely with the task setup proposed in Nguyen et al.~\cite{nguyen2025}, we consider a mixture consisting of a total of $M=5$ active sources, comprising $3$ targets and $2$ interfering sources. Note, however, that the metric proposed here applies to more general scenarios.

\subsection{Classical SDR}
\label{sec:classicalsdr}

In its most basic form, the SDR~\cite{LeRoux2019sisdr} between a target signal $s$ and its estimate $\hat{s}$ is defined as:
\begin{equation}
\label{eq:SDR}
    \text{SDR}(\hat{s},s) = 10 \log_{10} \left( \dfrac{\parallel s \parallel ^2}{\parallel \hat{s} - s \parallel ^2}\right).
\end{equation}
In the context of source separation, it is customary to compute a permutation-invariant variant of the SDR as follows:
\begin{equation}
\label{eq:classicalSDR}
    \dfrac{1}{N}\sum_{i=1}^{N} \text{SDR}(*\hat{s}_i, s_i),
\end{equation}
where ($\hat{s}_1, \dots, \hat{s}_N$) is the set of estimated sources, and $*\hat{s}_i$ is the permuted estimated source with the highest similarity with~$s_i$, i.e., such that this set of permuted sources yields the highest possible total SDR. We refer to \eqref{eq:SDR} as the SDR and to ~\eqref{eq:classicalSDR} as the \emph{classical} SDR.

\subsection{CA-SDR}
\label{sec:casdr}

The classical SDR defined above is straightforward to compute, but it does not take the class labels into account. Thus, the CA-SDR was proposed to additionally evaluate sound event classification~\cite{nguyen2025,yasuda2025}. Let 
$\hat{C} = (\hat{c}_1,\dots,\hat{c}_N)$ denote the predicted class labels associated with ($\hat{s}_1, \dots, \hat{s}_N$). Then, the CA-SDR is defined as follows:
\begin{equation}
\label{eq:ca-sdr}
   \text{CA-SDR} = \dfrac{1}{\mathcal{N}}\sum\limits_{k \in \mathcal{K}} \textrm{SDR}(\hat{s}_k, s_k) ,
\end{equation}
where $\mathcal{K}$ is the set of source indices $k$ such that $c_k \in C \cap \hat{C}$, and $\mathcal{N}$ is a constant and depends on the aggregation strategy being used, as will be explained in Section \ref{sec:agg_strategy}.

Using the above definition, the CA-SDR evaluates the S5 systems on both the separation and classification tasks as follows: 
\begin{itemize}
    \item Identifying the sources corresponding to the TPs, FPs, and FNs by comparing the set of predicted labels and the set of reference labels.
    \item Computing the improvement in SDR only for the sources that were correctly classified (TPs), and assigning a $0$ dB value for the sources that were not classified correctly (FNs and FPs).
    \item Taking the mean of the above improvement values as per the aggregation strategy.
\end{itemize}
%

\subsection{Limitations of CA-SDR}
\label{sec:casdrlimits}

Here, we present possible limitations of the CA-SDR. This metric is computed based on a set of estimated sources, obtained by allocating each source to a class using the S5 system. This approach considers the set of classes that are predicted over all the estimated sources to compute classification errors, and assigns an estimated source to a matching target source of the same class. In principle, CA-SDR uses predicted labels for source matching. This overlooks cases where sources are correctly separated but allocated an incorrect or no label. 

A first case described in Table~\ref{tab:limitations_1} consists of two sources being allocated classes that are present in the original mixture (e.g., ``cough" and ``pour"), while the third source is not allocated any class. Moreover, the system assigns the label ``pour" to the source that was actually containing ``dishes". The CA-SDR matching will associate target and estimated sources based on their classes, so there will be two TPs for ``cough" and ``pour," and the third source is considered an FN as if ``dishes" was not detected. Hence, the SDRs are computed for the sources labeled as ``cough" and ``pour" resulting in addition of the term $\text{SDR}(*\hat{s}_2,s_3)$ to the metric calculation and leading to an overall low value. This low metric value could be interpreted as either bad separation or a classification error. This mis-classification becomes explicit (\textit{cf}. the rightmost column) by matching the sources based on classical SDR~\eqref{eq:classicalSDR}. The source that was assigned the label ``pour" is then denoted $*\hat{s}_2$ as it best matches the target source $s_2$.

A similar behavior is observed in a second case described in Table~\ref{tab:limitations_2}, where all sources are allocated one of the labels that were present in the original mixture, but two sources' labels are swapped (``dishes" and ``pour"). Here, there are now two FNs, two FPs, and a TP, while the CA-SDR calculates the results as if there were three TPs. The SDR relative to the sources classified as ``cough" i.e., $\text{SDR}(*\hat{s}_2, s_1)$ and as ``dishes" i.e., $\text{SDR}(*\hat{s}_1, s_2)$ will be negative as the estimated sources are not matched with the correct target sources. This again could be the sign of a poor separation or an incorrect classification. When matching the sources based on the optimal permutation in terms of SDR, this source swapping becomes explicit (\textit{cf}. the rightmost column).

In both cases, classification errors can result in low CA-SDR values. However, this is ambiguous, as a low CA-SDR value could also reflect properly classified sources with poor separation.

\begin{table}[t]
\begin{center}
	\caption{CA-SDR and classical SDR based sources matching when one estimated source is not allocated the proper label ($*\hat{s}_2$, cough) and one estimated source is not allocated any label ($*\hat{s}_3$, none). The boldface font indicates which of the terms (separated source or predicated classes) is used for the source matching.}
	\label{tab:limitations_1}
	\begin{tabular}{|c|c|c|c|c|}
		\hline
		
	\multirow{2}{*}{References}&\multicolumn{4}{c|}{Predictions matching}\\&\multicolumn{2}{c|}{CA-SDR}&\multicolumn{2}{c|}{Classical SDR}\\

		\hline
		($s_1$, cough)  & ($*\hat{s}_1$, \textbf{cough})    & TP  &($\textbf{*}\hat{\textbf{s}}_\textbf{1}$, cough)    & TP \\
		($s_2$, dishes) & ($*\hat{s}_3$, none) &FN  & ($\textbf{*}\hat{\textbf{s}}_\textbf{2}$, pour)     &FN+FP\\
		($s_3$, pour) & ($*\hat{s}_2$, \textbf{pour})     &TP &($\textbf{*}\hat{\textbf{s}}_\textbf{3}$, none) &FN    \\
		\hline
	\end{tabular}
    \end{center}
\end{table}

\begin{table}[t]
	\begin{center}
	\caption{CA-SDR and optimal SDR based sources matching when one estimated source is not allocated the proper label ($*\hat{s}_1$, dishes) and ($*\hat{s}_2$, cough).}
	\label{tab:limitations_2}
	\begin{tabular}{|c|c|c|c|c|}
		\hline
		
	\multirow{2}{*}{References}&\multicolumn{4}{c|}{Predictions matching}\\&\multicolumn{2}{c|}{CA-SDR}&\multicolumn{2}{c|}{Classical SDR}\\

		\hline
		($s_1$, cough)  & ($*\hat{s}_2$, \textbf{cough})    & TP  &($\textbf{*}\hat{\textbf{s}}_\textbf{1}$, dishes)    & FN+FP \\
		($s_2$, dishes) & ($*\hat{s}_1$, \textbf{dishes}) &TP  & ($\textbf{*}\hat{\textbf{s}}_\textbf{2}$, cough)     &FN+FP\\
		($s_3$, pour) & ($*\hat{s}_3$, \textbf{pour})     &TP &($\textbf{*}\hat{\textbf{s}}_\textbf{3}$, pour) &TP    \\
		\hline
	\end{tabular}
    \end{center}
\end{table}


\subsection{CASA-SDR}
\label{sec:casasdr}

In this section, we present CASA-SDR,\footnote{https://github.com/mishramayank1903/Metric-analysis-for-S5} and discuss how it alleviates the limitations of CA-SDR. 

CASA-SDR uses the following procedure for S5 evaluation:
\begin{itemize}
    \item Instead of first calculating the labeling errors on the predicted labels, we calculate the SDR between all possible pairs of reference and estimated sources and keep the best permutation, i.e., we utilize the classical SDR defined in~\eqref{eq:classicalSDR}.
    \item  Applying the same permutation to the class labels, ensuring that (source, label) tuples remain as predicted by the S5 system before the permutation phase. 
    \item Calculating the classification errors, i.e., TPs, FPs, and FNs, given the permutation of the sources obtained above. 
    \item Only keeping the score for sources with correct labels (TPs), and assigning a score of 0~dB to sources with incorrect labels.
    \item Computing the mean of improvement values as per the aggregation strategy.
\end{itemize} 
%
Formally, the CASA-SDR is defined as:
\begin{equation}
\label{eq:casa-sdr}
   \text{CASA-SDR} = \dfrac{1}{\mathcal{N}}\sum\limits_{k \in \mathcal{K}} \textrm{SDR}(*\hat{s}_k, s_k) ,
\end{equation}
which is very similar to~\eqref{eq:ca-sdr}, except it uses $*\hat{s}_k$ instead of $\hat{s}_k$. This modification allows for controlling the impact of the FPs/FNs on the metric, while in CA-SDR, these errors can reflect in either a $0$ dB or a low metric value. 
In essence, when there are labeling errors, e.g., as described in Tables~\ref{tab:limitations_1}, \ref{tab:limitations_2}, CASA-SDR avoids assigning a low or sometimes negative metric value that could be ambiguous, as it always takes the best permutation of sources. Therefore, with CASA-SDR, a lower value indicates a poor separation, while classification errors are explicit. 

\subsection{Aggregation strategies}
\label{sec:agg_strategy}

As described in the previous sections, both CA- and CASA-SDR are computed by assigning a 0~dB score to incorrectly classified sources. One way to further control the impact of classification errors is to consider a non-zero penalty. However, in this paper, we follow the setup used in Yasuda et al.~\cite{yasuda2025} and the penalties are not used. Instead, we adjust the constant $\mathcal{N}$ in~\eqref{eq:ca-sdr} and~\eqref{eq:casa-sdr}, via the following two \textit{aggregation strategies}:
\begin{itemize}
    \item Error-based (EB) aggregation: we set $\mathcal{N}$ as the sum of the number of TPs, FPs, and FNs. When using this aggregation method, the systems are penalized further based on the number of classification errors they made.
    \item Source-based (SB) aggregation: we set $\mathcal{N}$ as the number of reference sources. When using this aggregation method, the systems are not penalized further based on the number of classification errors.
\end{itemize}
In this paper, the CA-SDR default aggregation strategy is EB, following proposed in its original paper~\cite{nguyen2025}. However, for CASA-SDR, the default strategy is SB, which avoids over-penalizing the systems due to classification errors, as we already assign a $0$ dB score to FP and FN sources. These default aggregation strategies are used while comparing the metrics, unless stated otherwise, and their specific impact will be discussed in~\ref{sec:impact_of_agg_strategies}.

\section{Experimental protocol}
\label{sec:exp_protocol}

In order to study the behavior of the metrics, we conduct experiments in two ways. First, we conduct experiments under controlled conditions, where we systematically introduce errors to audio signals. Secondly, we compare our metrics on the outputs of real-world S5 systems that were submitted to Task 4 of the DCASE 2025 challenge.

\subsection{Datasets}
\label{sec:data}

In order to evaluate CA-SDR and CASA-SDR in a controlled setup, we design a custom dataset as follows. We use the Spatial Scaper library~\cite{Romn2024SpatialSA} to generate audio mixtures using a subset of the FSD 50k dataset~\cite{fonseca2022FSD50K}. We generate $10$ second-long audio mixtures using room impulse responses (RIRs) available in the Spatial Scaper toolbox. We crop the RIRs at $50$ ms in order to discard the late reverberations, such that the setup is close to that proposed by Yasuda et al.~\cite{yasuda2025}, who evaluated dry sources. We then combine these audio sources from distinct target classes to generate audio mixtures. We generate $500$ audio mixtures with a total of $5$ sources, out of which there were $3$ target sources.

For the second experimental scenario, i.e., the one that considers real-world S5 systems, we use the official data used in the corresponding challenge's task.

\subsection{Experiments in controlled conditions}
\subsubsection{Impact of classification errors}
\label{sec:classif_errors}

First, we focus on the impact of classification errors on the metrics' behavior. We analyze the metrics in an oracle source separation setup, i.e., without cross-contamination between the separated sources. We introduce labeling errors of three types on the predicted sources: deletion, substitution, and swapping, as illustrated in Table~\ref{tab:labeling_errors}. We calculate the classical SDR, CA-SDR, and CASA-SDR after the addition of white noise at a $10$~dB signal-to-noise ratio (SNR).

\begin{table}[t]
\begin{center}
    \caption{Classification error types on an example.}

	\label{tab:labeling_errors}
	\begin{tabular}{|c|c|c|c|}
    \hline
     References & Type  1   &  Type 2  & Type 3  \\
           &  (deletion)  &  (substitution) & (swapping) \\
       \hline
     cough  & cough     & cough     & cough  \\
     dishes & none    & telephone  & pour \\
     pour  & pour     & pour     & dishes \\
       \hline
	\end{tabular}
\end{center}    
    
\end{table}

\subsubsection{Impact of source cross-contamination}
\label{sec:impact_cross_cont}

To focus on the impact of cross-contamination between the sources, we fix the predicted labels as correct (i.e., oracle classification). We progressively cross-contaminate the predicted sources by adjusting a scaling factor $\alpha \in [0,1]$ to control the amount of interference.

We considered different cases of cross-contamination, but for clarity here we focus on the following most informative case where two sources are gradually swapped as $\alpha$ increases from $0$ to $1$ and the third source is kept the same as the reference.
\begin{equation}
        \label{eq:cross_contam}
        \hat{u}_1 = (1-\alpha) \cdot u_1 + \alpha \cdot u_2 + \epsilon,    
\end{equation}
where $\epsilon$ is a $60$ dB additive white noise to avoid infinite values during calculations, and similarly for $\hat{u}_2$. 




\subsection{Experiments with real-world S5 systems}

\label{sec:experimentsT4}
When conducting experiments on the submitted systems, we employ different aggregation strategies for calculating the metrics, thereby highlighting the impact of each strategy. We also report the improvements in CA-SDR and CASA-SDR over the score computed on the mixture signal $x$, which is customary in such evaluation setups~\cite{nguyen2025,yasuda2025}.

For Task $4$ of the challenge, $8$ teams submitted $24$ systems. We begin our analysis by comparing the default versions of the metrics on the baseline plus $7$ other systems for which the output data was readily available. We use these systems for a first-hand comparison. Then, for a more detailed analysis, we focus on three systems: the best and worst performing ones, denoted morocutti\_CPJKU \cite{Morocutti_2025_t4} and zhang\_BUPT \cite{Zhang_2025_t4}, respectively, and the baseline system \cite{nguyen2025}. For more information on the rankings, please visit the results webpage \footnote{https://dcase.community/challenge2025/task-spatial-semantic-segmentation-of-sound-scenes-results}.

\section{Results and Discussion}
\label{sec:results_and_disc}

\subsection{Results in controlled conditions}  
\label{sec:results_oracle}

\subsubsection{Impact of classification errors}
\label{sec:impact_of_classiferrors}

In Table~\ref{tab:comparison classification}, we present a comparison between the classical SDR, CA-SDR, and CASA-SDR for different types of classification errors. We observe that for type 1 and 2 errors, both CA-SDR and CASA-SDR produce similar values since in these cases there are two TPs, and the metrics assign a 0~dB value to the other source. Type 3 error results in different values for the three metrics. This is due to the source matching process that is different in CA-SDR and CASA-SDR. On one hand, CASA-SDR assigns the two non-TP sources a $0$~dB and the TP source a $10$~dB score, producing an overall metric value of $3.33$~dB. On the other hand, CA-SDR assigns a value greater than $0$ dB to the TP source and values lower than $0$ dB to the two swapped sources, resulting in an overall negative metric value.

This experiment shows that classical SDR is unsuitable for the S5 task because it yields identical scores regardless of the type of classification error. CA‑SDR, in contrast, strongly penalizes type 3 errors relative to types 1 and 2 but offers no explicit control over the impact of type 3 errors on the metric. CASA‑SDR does not distinguish between type 1 and type 2 errors, yet it allows controlling the influence of type 3 errors by assigning 0 dB to swapped sources, making it more informative than both classical SDR and CA‑SDR in this respect.

\begin{table}[t]
\begin{center}
 \caption{Metric values (in dB) at 10 dB SNR.}
	\label{tab:comparison classification}
	\begin{tabular}{|c|c|c|c|}
            \hline
               Metric  & Type  1   &  Type 2  & Type 3  \\
              &  (deletion)  &  (substitution) & (swapping) \\
               \hline
            Classical SDR  & 10.00     & 10.00     & 10.00  \\
            CA-SDR & 6.67    & 6.67  & -0.68  \\
           CASA-SDR & 6.67   & 6.67    & 3.33 \\
               \hline
    \end{tabular}
\end{center}    
\end{table}

\subsubsection{Impact of source cross-contamination}
\label{sec:impact_of_cross_cont}

\begin{figure}[t]
\centerline{\includegraphics[width=.68\linewidth]{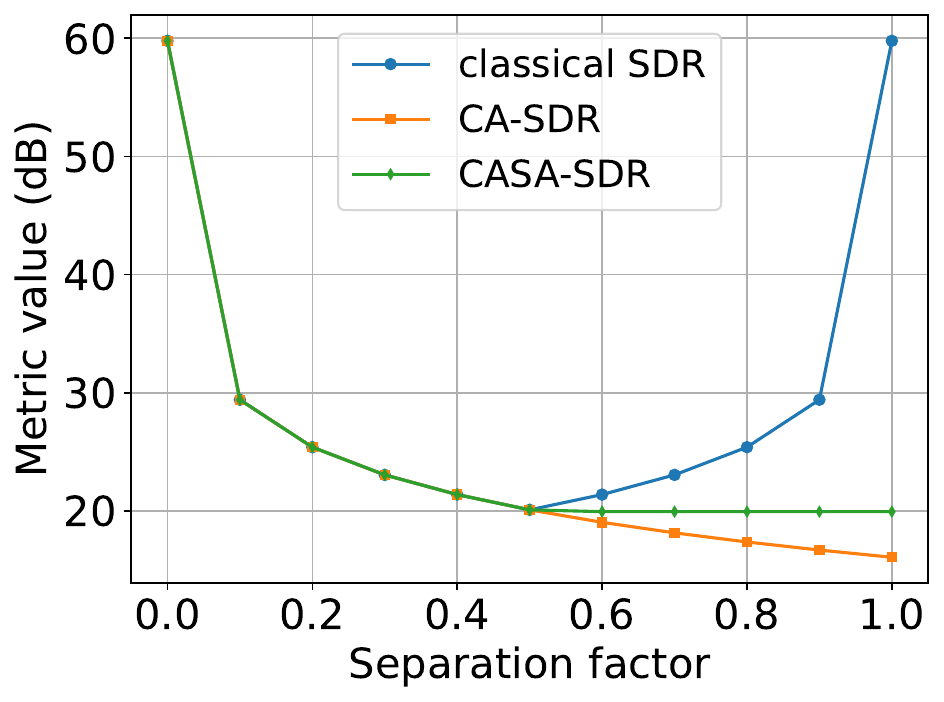}}
    \caption{Comparison of metrics in case of cross-contamination.}
  \label{fig:impact_of_cross_contamination}
\end{figure}

We now analyze how the metrics behave under cross-contamination between the sources, such that two of them are gradually swapped as described in Section~\ref{sec:impact_cross_cont}. The results are displayed in Figure~\ref{fig:impact_of_cross_contamination}. While all metrics decrease similarly as $\alpha$ increases, they behave differently for $\alpha>0.5$. When $\alpha>0.5$, the classical SDR increases because the permutation matches the swapped sources with the correct target source. CA-SDR is decreasing as the value between the swapped source and the target source is decreasing, but because of the source matching process, CA-SDR does not detect any classification error. In CASA-SDR, on the other hand, the swapped sources are correctly matched thanks to the permutation, but the metric detects the classification errors and sets a $0$~dB value for the corresponding sources.
Overall, this experiment reveals that CASA-SDR allows for identifying label swapping when such an error occurs because of poor source separation, while this is not possible with CA-SDR.  

\subsection{Results with real-world S5 systems}

\begin{figure}[t]
  \centering
  \includegraphics[width=\linewidth]{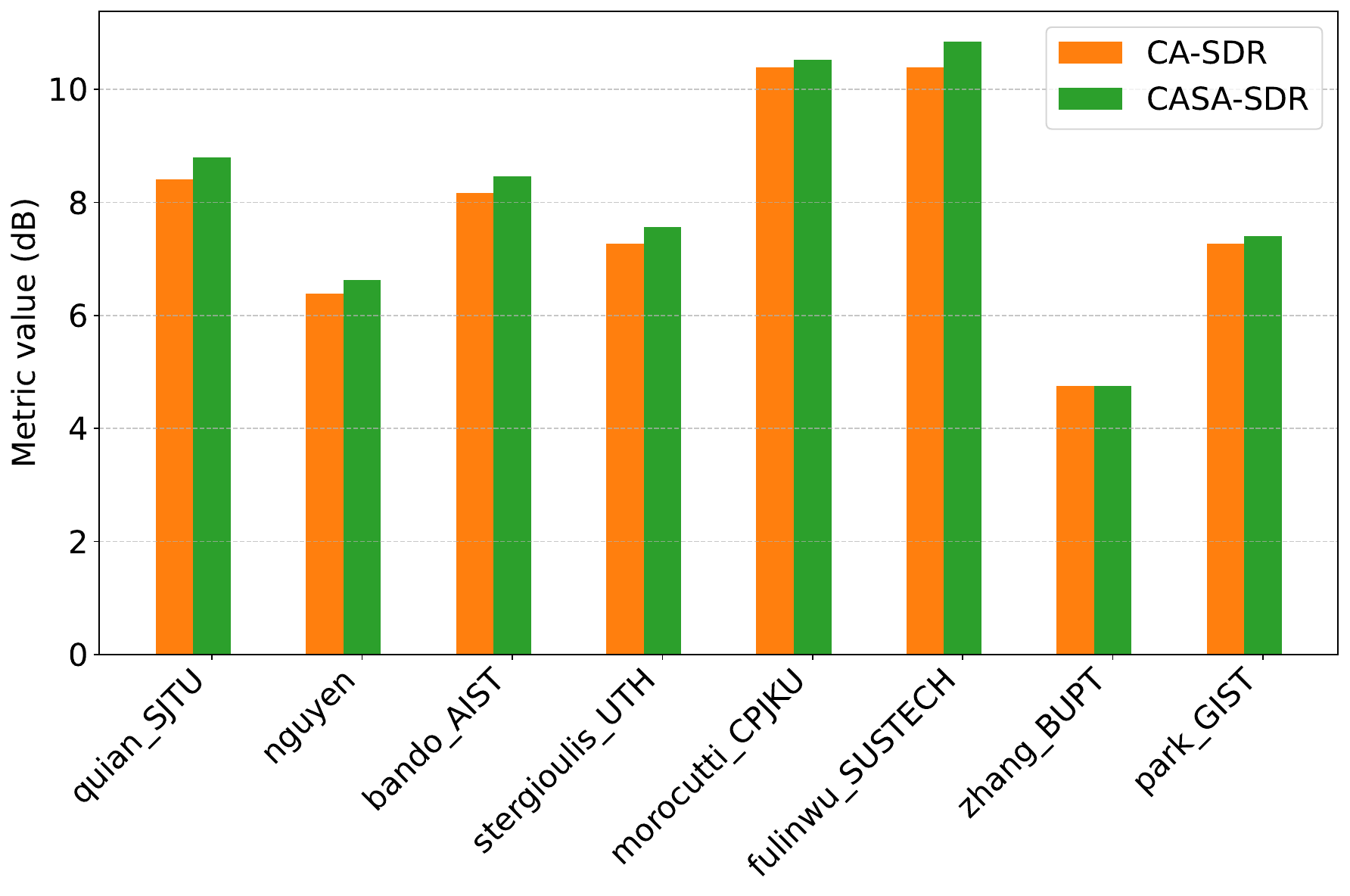}
    \caption{Comparison of metrics with default aggregation strategy over systems.}
  \label{fig:default_aggregation}
\end{figure}

We now study the metrics using the submissions to the DCASE challenge. Note that since the classical SDR does not yield any information about the classification performance of the systems, we do not report it in what follows, and we focus on CA-SDR and CASA-SDR. From the results displayed in Figure~\ref{fig:default_aggregation}, we observe that CA-SDR over-penalizes the systems compared to CASA-SDR, resulting in a lower metric value. To further reveal the impact of differences between CA-SDR and CASA-SDR, the rest of our analysis focuses on three systems, following the setup described in Section~\ref{sec:experimentsT4}.

\subsubsection{Aggregation strategies}
\label{sec:impact_of_agg_strategies}

First, we study the impact of aggregation strategies onto metrics. The results are presented in Table~\ref{tab: aggregation_comparison}. We observe that the difference between CA-SDR and CASA-SDR when using the same aggregation strategy is at most of order $10^{-2}$. This difference is explained by the number of classification errors counted by both metrics. While for CASA-SDR all these errors could potentially be swaps, for CA-SDR, there cannot be only swaps, as this would result in an overall lower metric value when compared to CASA-SDR, as explained in \ref{sec:impact_of_classiferrors}, but this is not the case here. Hence, we investigate in more details the number of classification errors counted by each metric.

\begin{table}[t]
\begin{center}
    \caption{CA-SDR vs CASA-SDR on different aggregation strategies.}
	\label{tab: aggregation_comparison}
\begin{tabular}{|c|cc|cc|}
  \hline
  System & \multicolumn{2}{c|}{CA-SDR} & \multicolumn{2}{c|}{CASA-SDR} \\
\cline{2-5}
         & EB & SB & EB & SB \\
  \hline
  morocutti\_CPJKU & 10.39 & 10.55 & 10.37 & 10.53 \\
  zhang\_BUPT      &  4.75 & 4.76 & 4.75 & 4.76 \\
  baseline         &  6.38 & 6.67 & 6.38 & 6.62 \\
  \hline
\end{tabular}
\end{center}
\end{table}

\subsubsection{Number of classification errors}

Table~\ref{tab: numberof_tp_fn_fp} presents the number of TPs, FPs, and FNs predicted by the systems when using the two metrics. As we can see for all systems, CASA-SDR predicts a smaller number of TPs than CA-SDR. This is in line with the fact that CA-SDR allows sources with poor separation quality to be identified as TPs as long as the corresponding label is correct, a case similar to Table~\ref{tab:limitations_2}. We also observe that the reduction in TPs obtained with CASA-SDR, relative to CA-SDR, is exactly matched by a corresponding increase in FPs and FNs for CASA-SDR. This, in turn, could indicate potential swap cases, and hence we investigate this in the next experiment.

\begin{table}[t]
	\begin{center}
    \caption{Number of TPs, FNs, and FPs per system.}
	\label{tab: numberof_tp_fn_fp}
\begin{tabular}{|c|ccc|ccc|}
  \hline
  System & \multicolumn{3}{c|}{CA-SDR} & \multicolumn{3}{c|}{CASA-SDR} \\
  \cline{2-7}
         & TPs & FPs & FNs & TPs & FPs & FNs \\
  \hline
  morocutti\_CPJKU & 2527 & 200 & 713 & 2504 & 223 & 736 \\
  zhang\_BUPT     &  1278 & 418 & 1962 &  1196 & 500 & 2044 \\
  baseline     &  2364 & 402 & 876 &  2283 & 483 & 957 \\
  \hline
\end{tabular}
\end{center}
\end{table}


\subsubsection{Identification of error scenarios}

In Section~\ref{sec:casdrlimits}, we discussed a potential limitation of CA-SDR regarding identifying swap cases. This was further confirmed experimentally in Section~\ref{sec:results_oracle}. In this section, we aim to investigate this behavior using the submitted systems and official challenge data. From Table~\ref{tab: numberof_tp_fn_fp}, we can see that for the morocutti\_CPJKU, zhang\_BUPT, and baseline systems, $23$, $82$, and $81$ TPs predicted by CA-SDR are not predicted as TPs by CASA-SDR, respectively. On average, the CA-SDR values for the morocutti\_CPJKU, zhang\_BUPT, and baseline systems for these TPs are $1.07$, $-2.71$, and $-0.01$ dB, respectively. These CA-SDR values are extremely low or negative, while for CASA-SDR, these values are $0$ dB by design. This led to a closer investigation of these TPs, and we found that when the overall metric value for CA-SDR is negative, then there is definitely a swap case. In these swap cases, the reason for a swap can be either because of label swapping as explained in Section~\ref{sec:classif_errors}, or because of cross-contamination between sources as explained in Section~\ref{sec:impact_cross_cont}. For the cases where the overall values of CA-SDR are positive but still less than $5$ dB, it is difficult to draw such conclusions.

Finally, we inspected the labels of the sources in the case of swapping, and found that these occur mostly when the mixtures have two or more sources that are similar in terms of audio content, e.g., ``hair dryer" and ``vacuum cleaner", ``doorbell" and ``musical keyboard", etc. This finding will need a more in-depth investigation to be fully confirmed.

\section{Conclusion}
\label{sec:conc}

In this paper, we analyzed the CA-SDR metric for evaluating S5. We highlighted a potential limitation of this metric. We proposed an alternative CASA-SDR metric, which refines the identification of separation and classification errors. The proposed framework offers an alternative evaluation methodology, shifting from a classification-centric approach to one that is separation-centric. If the target application for S5 does not require high separation quality and is mainly concerned with label accuracy, CA-SDR is an appropriate choice. It can effectively evaluate systems for tasks where source matching is the main goal, and high separation quality is not essential.
However, if the application demands both accurate labeling and high separation quality, CASA-SDR is a viable option. This metric performs optimal source matching and penalizes systems that produce poor separation results.
A further investigation would be required into the system architectures to assess their behaviors in cases where the overall metric values are low but positive, and also into which class labels are perceptually similar, leading to a much harder classification. Besides, it would also be interesting to work on modifying the metrics by adding non-zero penalties, such as the output classical SDR values between the misclassified sources, and then compare various systems using those improved versions.


\newpage
\section{Acknowledgments}
This research was carried out with the support of the French National Research Agency as part of the CONFLUENCE project number \text{ANR-23-EDIA-0003}. Experiments presented in this paper were carried out using the \href{https://www.grid5000.fr}{Grid'5000}
testbed, supported by a scientific interest group hosted by Inria and including CNRS, RENATER, and several universities as well as other organizations.

\bibliographystyle{IEEEbib}
\bibliography{refs}

\end{document}